\documentstyle[epsfig]{article}

\renewcommand{\thefootnote}{\fnsymbol{footnote}}

\begin{document}
\begin{flushright}
SLAC-PUB-8669 \\
hep-ph/0010182 \\
October 2000
\end{flushright}
\bigskip
\begin{center}
\Large\bf
$\gamma^*\gamma\to\pi\pi$ at large $Q^2$ \footnote{Talk presented
at PHOTON 2000, Ambleside, England, August 2000. To appear in the
proceedings.}
\end{center}
\bigskip
\centerline{Markus Diehl\,$^1$\footnote{Work supported by Department of
Energy contract DE--AC03--76SF00515 and by the Feodor Lynen Program of
the Alexander von Humboldt Foundation.}\, ,
Thierry Gousset\,$^2$ and Bernard Pire\,$^3$}
\smallskip
\begin{center}
1. Stanford Linear Accelerator Center, Stanford University,\\
Stanford, CA 94309, U.S.A.\\
2. SUBATECH, B.P.~20722, 44307 Nantes, France \footnote{Unit\'e
mixte 6457 de l'Universit\'e de Nantes, de l'\'Ecole des Mines de
Nantes et de l'IN2P3/CNRS.}\\ 
3. CPhT, Ecole Polytechnique, 91128 Palaiseau, France
\footnote{Unit\'e mixte 7644 du CNRS.}
\end{center}
\bigskip
\centerline{\large\bf Abstract}
\begin{center}
\parbox{0.9\textwidth}{The QCD analysis of the process $\gamma^*
\gamma \to \pi \pi$ at large $Q^2$ and small center-of-mass energy
allows one to access a new hadronic observable describing the
exclusive transition from a $q \bar q$ or $gg$ state to a pair of
mesons. A fruitful study may be envisaged at existing machines.}
\end{center}
\bigskip
\setcounter{footnote}{0}
\renewcommand{\thefootnote}{\arabic{footnote}}

\section{Introduction}

Exclusive hadron production in two-photon collisions provides a tool
to study a variety of fundamental aspects of QCD and has long been a
subject of great interest (cf.\ \cite{Terazawa,Bud,photon-conf}
and references therein). Recently a new aspect of this has been pointed
out, namely the physics of the process $\gamma^* \gamma \to \pi \pi$
in the region where $Q^2$ is large but $W^2$
small~\cite{DGPT,Long}. This process factorizes~\cite{MRG,Freund} into
a perturbatively calculable, short-distance dominated scattering
$\gamma^* \gamma \to q \bar q$ or $\gamma^* \gamma \to g g$, and
non-perturbative matrix elements measuring the transitions $q \bar q
\to \pi \pi$ and $g g \to \pi \pi$. We call these matrix
elements generalized distribution amplitudes (GDAs) to emphasize their
close connection to the distribution amplitudes introduced long
ago in the QCD description of exclusive hard
processes~\cite{LepageBrodsky}.

\section{Factorization}
\label{sec:fa}

\begin{figure}[b!]
\centerline{\epsfig{file=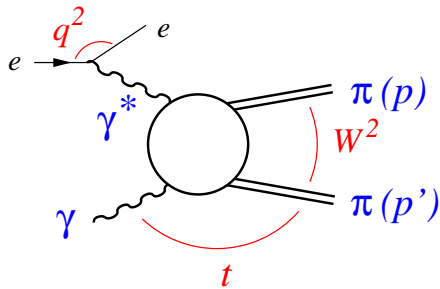,width=0.5\textwidth}}
\vspace{10pt}
\caption{The subprocess $\gamma^*\gamma\to\pi\pi$ in the reaction
$e+\gamma\to e+\pi+\pi$.}
\label{fig:ggtopp}
\end{figure}

We are interested in $e+\gamma\to e+\pi\pi$. To lowest order in QED
the interaction between the lepton and the hadron side is mediated
by one-photon exchange. The contribution we focus on is the one where
the $\gamma^*\gamma\to\pi\pi$ subprocess appears, as depicted in
Fig.~\ref{fig:ggtopp}. The two diagrams where both photons attach to
the lepton line are referred to as the bremsstrahlung
contribution. We notice that in this case the pions are in a C-odd
state, in contrast to the former one where they emerge in a C-even
state. The phenomenological interest of this observation will be
discussed in Sect.~\ref{sec:ph}.

The kinematical regime in which we study the reaction is that of large
$Q^2=-q^2$, where $q=k-k'$ is the momentum tranferred from the
electron (see Fig.~\ref{fig:ggtopp}), and small $W^2=(p+p')^2$. To be
specific we shall explore the domain $Q^2\ge 4$~GeV$^2$ and $W^2\le
1$~GeV$^2$. The $Q^2$ at which the leading contribution to be
discussed starts to drive the cross section is, however, a matter of
experimental determination because present theory can at best estimate
the size of power corrections~\cite{KM}.

\begin{figure}[b!]
\centerline{\epsfig{file=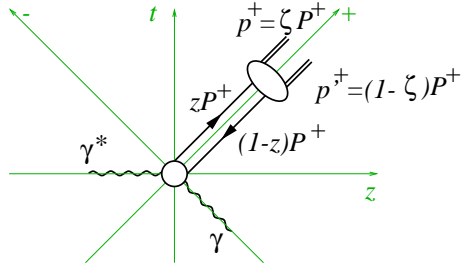,width=0.5\textwidth}}
\vspace{10pt}
\caption{Spacetime cartoon for $\gamma^*\gamma\to\pi\pi$ in the Breit
frame.}
\label{fig:ggtoppst}
\end{figure}

Let us now discuss the factorization of the process at large $Q^2$. To
do this it is useful to visualize the process in the Breit frame, where
the incoming real photon of momentum
$-\frac{1}{2}(Q+W^2/Q)\,\hat{z}$ collides with a static virtual photon
of momentum $Q\hat{z}$ and the pion pair emerges with momentum
$\frac{1}{2}(Q-W^2/Q)\,\hat{z}$. Neglecting $W^2$ compared with $Q^2$,
a simple spacetime cartoon can be drawn (see Fig.~\ref{fig:ggtoppst}),
which in addition shows the momentum sharing between the two pions
on one hand, and between the two partons on the other hand.

Although a complete proof of factorization relies on a very detailed
study of loop corrections, we may motivate it by emphasizing the
similarities between the one-pion and the two-pion channels. The
result of factorization is that the amplitude of the
$\gamma^*\gamma\to\pi\pi$ reaction is a combination of a short
distance transition from $\gamma^*\gamma$ to two partons (two quarks
or two gluons) and a long distance process of hadronization of two
partons into two hadrons.

\begin{figure}[b!]
\centerline{\epsfig{file=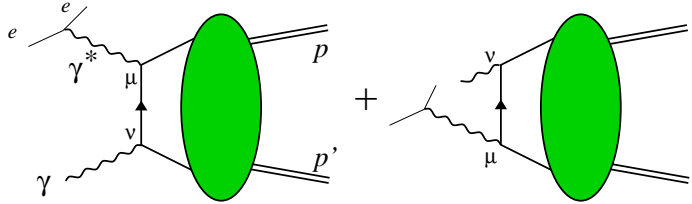,width=0.75\textwidth}}
\vspace{10pt}
\caption{Leading order amplitude.}
\label{fig:ggtopplo}
\end{figure}

At leading order in $\alpha_S$, the two photons couple to a $q\bar q$
pair. The relevant diagrams are displayed in Fig.~\ref{fig:ggtopplo}.
The matrix element of the time ordered product of the two
electromagnetic current reads
\begin{equation}\label{eq:am}
T^{\mu\nu}=-g_T^{\mu\nu}\sum_q \frac{e_q^2}{2}
\int_0^1 dz\,\frac{2z-1}{z(1-z)}\,\Phi^+_q(z,\zeta,W^2),
\end{equation}
where $\Phi^+_q$ is the quark component of the two-pion distribution
amplitude. The index $+$ indicates that the process selects the C-even
component of the quark GDA. We notice that the amplitude (\ref{eq:am})
is independent of $Q^2$. This is understood to be true up to
logarithmic scaling violation. We also remark that both photons have
the same helicity. This property is specific to the $q\bar q$
channel. In the gluon channel, i.e., for the $\gamma^*\gamma\to gg$
subprocess, the photons can have opposite
helicities~\cite{Kivel:1999sd},
but still the virtual photon has to be transversely polarized. The
phenomenological consequences of all these facts will be discussed in
Sect.~\ref{sec:ph}.

\section{The generalized distribution amplitude}
\label{sec:ge}

The operator definition of the two-pion distribution amplitude is, in
$A^+=0$ gauge,
\begin{eqnarray*}
\Phi_q(z,\zeta,W^2)&=&\int\frac{dx^-}{2\pi}e^{-iz(P^+x^-)}
\langle\pi(p)\pi(p')|\bar q(x^-)\gamma^+ q(0)|0\rangle,\\ 
\Phi_g(z,\zeta,W^2)&=&\frac{1}{P^+}\int\frac{dx^-}{2\pi}e^{-iz(P^+x^-)}
\langle\pi(p)\pi(p')|F^{+\mu}(x^-){F_\mu}^+(0)|0\rangle.
\end{eqnarray*}
The standard analysis of QCD radiative corrections shows that
the amplitude (\ref{eq:am}) is modified by logarithms of $Q^2$. The
leading logarithmic corrections may be factorized into the
distribution 
amplitude, which thus depends on a factorization scale $\mu^2$ through
a linear combination of terms $(\log \mu^2)^{-d_n}$ with anomalous 
dimensions $d_n$. We refer the reader to Ref.~\cite{Long} for a review
in the present context. We only mention here that in the considered
channel quarks and gluons mix under evolution and that all $d_n$'s are
positive, except for one which is zero. This implies that the GDA
tends to a 
non vanishing asymptotic form when $\mu^2\to\infty$. This asymptotic
form reads:
\begin{eqnarray*}
\sum_{q=1}^{n_f} \Phi_q^+(z,\zeta,W^2)
&=&18 n_f z(1-z)(2z-1)\\
&&\times\left[ B_{10}(W^2)+B_{12}(W^2)\, P_2(2\zeta-1) \right],\\
\\
\phantom{\sum_{q}} \Phi_g(z,\zeta,W^2)&=&48z^2(1-z)^2\\
&&
\times\left[ B_{10}(W^2)+B_{12}(W^2)\, P_2(2\zeta-1) \right].
\end{eqnarray*}
One finds that the first subasymptotic term has the same
polynomials dependence in $z$ and $\zeta$, whereas solutions with yet
higher $d_n$ involve polynomials of higher degree in both $z$ and
$\zeta$.

One interesting phenomenological consequence of the interrelation of
$z$ and $\zeta$ is that, since the $\zeta$ dependence may be
rewritten as a partial wave expansion, the $\cos\theta$ dependence of
the cross section provides information on the $z$ behavior of
$\Phi$. The latter would be otherwise difficult to extract since the
amplitude is given as an integral over this variable (see
Eq.~(\ref{eq:am})).

Let us now construct a simple model for the GDA. To perform this
task, we take advantage of the energy-momentum sum-rules:
\begin{eqnarray*}
\int_0^1 dz\,(2z-1)\,\Phi_q(z,\zeta,W^2)&=&
\frac{2}{(P^+)^2}\langle\pi^+(p)\pi^-(p')|T_q^{++}(0)|0\rangle,\\
\int_0^1 dz\,\Phi_g(z,\zeta,W^2)&=&
\frac{1}{(P^+)^2}\langle\pi^+(p)\pi^-(p')|T_g^{++}(0)|0\rangle,
\end{eqnarray*}
\noindent
and of the  form factor decomposition of $T^{\mu\nu}$:
\begin{eqnarray*}
\langle\pi(p)\pi(p')|T_{q,g}^{\mu\nu}(0)|0\rangle
&=&\frac{1}{2}
T^{(1)}_{q,g}(W^2)\,[(p+p')^\mu(p+p') ^\nu-W^2g^{\mu\nu}]
\\
&+&\frac{1}{2} T^{(2)}_{q,g}(W^2)\,(p-p')^\mu (p-p')^\nu.
\end{eqnarray*}
We have to perform an analytic continuation from $W^2$ to $0$, which
leads to
\[
\frac{9n_f}{10}\,B_{12}(0)=\sum_q T^{(2)}_{q}(0)=R_\pi
\]
where $R_\pi\approx 50\%$ is the total momentum fraction carried by
quarks and antiquarks in a pion.

We simplify the discussion on energy dependence by using first that
below the inelastic threshold the partial wave phases are related to
the phase shifts $\delta_0(W^2)$ and
$\delta_2(W^2)$ of $\pi\pi$ elastic scattering as a consequence of
Watson's theorem\cite{Pol}. We then assume $|B_{12}(W^2)|$ to be
constant and thus equal to 
$B_{12}(0)$. Finally, we estimate $B_{10}$ through a soft pion theorem
that gives $B_{10}(0)=-B_{12}(0)$~\cite{Pol}.

Our model two-pion distribution amplitude thus reads
$$
\Phi_{u/d}=10z(1-z)(2z-1)R_\pi
\left[\frac{\beta^2-3}{2}\,e^{i\delta_0(W^2)}
+\beta^2\,e^{i\delta_2(W^2)}\,P_2(\cos\theta)\right]
$$

\section{Phenomenology}
\label{sec:ph}

Let us now briefly outline some useful phenomenological features of
the $e\gamma\to e\pi \pi$ process (see Ref.~\cite{Long} for more
details).

There are three independent helicity amplitudes $A_{++}$, $A_{0+}$ and
$A_{-+}$, but the first one dominates at large $Q^2$ and may be
written as
\[
A_{++} = \sum_q \frac{e_q^2}{2}\int_0^1\! dz\,\frac{2z-1}{z(1-z)}\,
\Phi^{\pi\pi}_q(z,\zeta,W^2),
\]
at leading order in $\alpha_S$. The two other amplitudes are
non-leading:
\[
A_{0+}/ A_{++}\propto 1/Q,\quad\quad
A_{-+}/ A_{++}\propto\alpha_S(Q^2).
\]

In the case of  $ \pi^+\pi^-$ production, the cross section gets a
contribution from the Bremsstrahlung process and can be decomposed as
$$
d\sigma=d\sigma_B+d\sigma_I+d\sigma_G .
$$
The interference with the bremsstrahlung process allows us to access
the 
$\gamma^\ast\gamma$ process at the amplitude level, thanks to the
different C-conjugation properties of the two processes. This
contribution is selected by charge asymmetries such as
$$
d\sigma(\pi^+(p)\pi^-(p'))-d\sigma(\pi^-(p)\pi^+(p'))
$$
and by angular distributions derived from this.
This interference part may be written as
$$
d\sigma_{I}(Q^2,W^2,\cos\theta,\varphi)\propto
e_l\,\Big(C_0+C_1\cos\varphi+C_2\cos2\varphi+C_3\cos3\varphi\Big)
$$
with the dominant term at large $Q^2$ given by
\begin{eqnarray*}
C_1&=&{\mathrm{Re}}\Big\{F_\pi^* A_{++}\Big\}\,
      [1-(1-x)(1+\epsilon)]\sin\theta\\
&-&{\mathrm{Re}}\Big\{F_\pi^* A_{0+}\Big\}
\sqrt{2x(1-x)}\,2\epsilon\cos\theta\\
&+&{\mathrm{Re}}\Big\{F_\pi^* A_{-+}\Big\}(1-x)\sin\theta
\end{eqnarray*}

With our model distribution presented above we estimate
counting rates around $10^4$ events for an
integrated luminosity of 20--30~fb$^{-1}$ at an $e^+e^-$ c.m.\ energy
of 10 GeV. This would allow one to
measure the interference term with some $O(10\%)$ statistical errors.

\section{Summary}

The study of $\gamma^\ast \gamma\to\pi^+\pi^-$, $\pi^0\pi^0$ at
large photon virtuality and small c.m.\ energy is
a powerful new tool for investigating the confining mechanism which
takes place in the transformation of quarks or gluons into two
mesons. The simplest model for the two-pion distribution amplitude
uses the asymptotic
solution to the evolution equations, $\pi\pi$ elastic phase shifts,
and $R_\pi$.

Encouraging rates are predicted for the kinematics and luminosity of
existing $B$ factories, and one may expect data to come soon from
the BABAR, BELLE, and CLEO experiments.

The same GDA appears in deep electroproduction processes
too\cite{rho-pi-pi}, where it opens the possibility of extracting
skewed parton
distributions from $2 \pi$ electroproduction without careful
separation of resonant $\rho$ vs.\ non-resonant $2 \pi$ production.

\section*{Acknowledgments}

We acknowledge useful discussions with O.V. Teryaev.

\end{document}